\documentclass[preprint,aps,pre]{revtex4}
\usepackage{graphicx,color}
\begin{document}
%\draft
%\tightenlines
\title{\bf{ 
Driven diffusive system with non-local perturbations}}
\author{Sutapa Mukherji}
\affiliation{ Department of Physics, Indian Institute of
  Technology, Kanpur, India}
%\onecolumn
\date{\today}
\widetext
\begin{abstract}  
We investigate the impact of non-local perturbations on driven 
diffusive systems. Two different problems  are considered here.
In one case, we introduce a non-local particle conservation along
the direction of the drive and in another case, we incorporate
a long-range temporal correlation in the noise present in the 
 equation of motion. 
The effect of these perturbations on the anisotropy exponent or on the 
scaling  of the two-point correlation function is studied  
using renormalization group analysis. 
\end{abstract}
\pacs{ }

\maketitle

%\begin{multicols}{2}
\section{introduction}
Driven diffusive systems have a special role in the study of
non-equilibrium systems because of the unusual steady state properties
that are not found in equilibrium.  
The main aspects which are crucial for
these new non-equilibrium properties are particle conservation,
spatial anisotropy  associated with the drive and the existence of a
non-equilibrium steady state \cite{ziarev}.  A simple paradigmatic
model that captures all the above features is the so called KLS model
named after Katz, Lebowitz and Spohn \cite{katz,katz1}.  This model
describes diffusing lattice gas particles in contact with a thermal
bath.  In addition to the attractive inter-particle interaction, 
there is a driving force acting in a specific
direction in the $d$-dimensional space. The rate of hopping is biased
along
 the direction of the drive. The periodic boundary condition
along and transverse to the drive implies a toroidal geometry and in
this geometry even though the drive looks like a potential locally,
there is no global potential.  As a result, in the steady state there
is a steady flow of particle- and energy-current looping around the
toroid. The steady flow of energy results from the fact that the
particles gain energy from the drive and loose it in the thermal bath.

Since the appearance of the KLS model, various techniques such as mean
field theory, renormalization group (RG) analysis, numerical
simulations have been employed to investigate its steady state.
Although numerical simulations show the presence of the order-disorder
transition for all strength of the drive \cite{leung1}, the behaviors
above, below and at criticality are quite different from those in the
corresponding undriven case. Below criticality, apart from the
critical exponent $\beta$ or the critical temperature $T_c$, the most
crucial differences from the equilibrium Ising model appear through
the shape of the coexistence curve and the shape of the coexisting
regions which are strip-like instead of droplets \cite{valles}.  Above
$T_c$, the most remarkable phenomenon is the existence of power-law
correlations \cite{katz1,zhang}. 
According to the  RG ideas in statistical 
mechanics, such scale invariant properties should be 
associated with RG fixed points. RG studies 
on  a system of diffusing, non-interacting
 particles, under drive \cite{jsdiff} indeed support this idea. 
This model is argued  as a special limit of the KLS model in which
the temperature and the drive strength are large with a fixed ratio of
the two \cite{bks}.
The RG analysis shows that for  $d>2$,  the high temperature properties
 of the standard model are controlled by a line of 
fluctuation-dissipation theorem (FDT) violating fixed 
points in contrast with  a single $T=\infty$ fixed point for 
the equilibrium Ising model. For $d<2$ there are both FDT restoring 
and FDT violating locally stable fixed points. 
 Striking differences appear in the
universality and scaling in the critical region that has been studied
in great detail.  The fixed point which governs the critical
properties of the undriven system becomes unstable and a new fixed
point determines the critical properties of the driven system
\cite{janssen}.  As a result, universal exponents are different and in
particular, there is an  anisotropy exponent
that distinguishes the directions along and perpendicular to the
drive. Although exponents are calculated using  $\epsilon(=5-d)$-expansion
scheme after one-loop RG analysis, they are exact due to the invariance of
the system under Galilean transformation.
                                                                             
In this paper, our focus is on the critical region and we are
interested in two problems that are related to the particle
conservation and the noise in the system (motivations discussed
below).  In the first problem, we relax the constraint of particle
conservation by introducing a non-locality and study its impact on the
KLS model.  In the second problem, the impact of a long-ranged
temporally correlated noise is studied because such temporal
correlations are expected to disturb the Galilean invariance.
 
Particle conservation is an important ingredient in driven diffusive 
systems.  However,
particle conservation need not necessarily be maintained locally and
it can be  satisfied in a  more global way.  
In  the lattice-gas language,  particles  may be 
interchanged over a long distance unlike the 
nearest-neighbour spin-exchange  Kawasaki-dynamics.
% the spin exchange is not restricted to
%the nearest neighbor only, but the exchange can take place with  a far
%away spin also.
 A few recent studies on  critical dynamics of 
model B\cite{bray,bray1}, model C
\cite{sen} and model D\cite{sen1} show that 
the non-locality in conservation results in drastic 
changes in the dynamical class.
In view of these recent observations and the importance of the 
conservation in 
driven diffusive systems, it is
natural to ask how crucial is the locality condition of the 
 conservation for the dynamics of the driven diffusive system.

 In driven diffusive systems, the noise is usually considered to have a
Gaussian distribution with a short-range  correlation 
in space and time. 
Though a long-range spatial correlation in the   noise 
in the high temperature version of the  KLS model  
 \cite{becker}  
exhibits interesting crossover behavior \cite{js} 
with the increase of  the range of the noise correlation, 
the effect of a  long-range temporal correlation
near the critical point need not be so.  
We have already pointed out that several
of the known results depend crucially on the Galilean invariance
\cite{janssen} ensuing from the delta-correlation of the noise in time.
However, this delta-correlation is not a necessity because the
FDT need not be respected in
non-equilibrium.  Also one notes that any temporally correlated noise
would break the Galilean invariance.  In fact,  long-ranged
temporal correlation in the noise is well-known in various problems that
include biological, physical and economical systems. Such temporally
correlated noise may also occur if the source is another correlated
system, which, e.g., can give rise to long-range time correlation at
criticality.  In our second problem, we study the effect of a temporally
correlated noise on the known results of driven diffusive systems described
by the KLS model.

Since our focus is on the long distance and large time behavior of
systems, a convenient starting point would be an appropriate continuum
model based on the Langevin equation of motion.  Our interest is in
the critical region where the dynamic RG analysis serves as an useful
machinery. We, therefore, adopt a momentum-shell RG technique in the 
following to
address the above questions. 

The rest of the paper is organized as follows. In the next section, we 
discuss the effect of  the 
 non-local conservation. After discussing the  model and 
already known results in certain  specific cases, 
we move on to the discussion of  our RG results in the next subsection. 
In section III, the effect of a  temporally correlated noise 
is discussed.                                      
We conclude by summarizing the results in section IV.

\section{ non-local particle conservation}
 The non-locality in the conservation is introduced through  a 
non-local kernel in the particle current density. The local problem 
which is the usual KLS model can be retrieved by considering
the  appropriate limit.

\subsection{Model and the known results}
Associated with  the conserved particle density $S({\bf x},t)$, there  is
a continuity equation 
\begin{eqnarray}
\partial_t S({\bf x},t)+\partial_\parallel J_\parallel+
{\bf \nabla}_\perp\cdot {\bf J}_\perp=0 \label{cons},
\end{eqnarray}
where the $J_\parallel$ and ${\bf J}_\perp$ are the current densities 
along and perpendicular to the drive.  In the absence of the drive, the 
deterministic parts of the  current densities can  be obtained from the 
gradient of the chemical potential. The total current densities, therefore, 
have the standard form                             
\begin{eqnarray}
        \ \ \ \ \ \ {\bf J}_\perp=-\lambda 
{\bf \nabla}_\perp \frac{\delta {\cal H}}{\delta 
S}+ {\bf \zeta}_\perp({\bf x},t) \ \ \ {\rm and} \label{perpcurrent}\\
        \ \ \ \ \ \  J_\parallel=\int d^d{ x'}\chi({\bf x}-{\bf x'})
[-\lambda \partial'_\parallel \frac{\delta {\cal 
H}}{\delta S({\bf x'})}+\zeta_\parallel({\bf x'},t)],\label{parallel}
\end{eqnarray}
where $\zeta_\perp$ and $\zeta_\parallel$ are the noisy parts of the 
current densities perpendicular and parallel to the drive respectively 
and $\lambda$ is the transport coefficient. Here and in the following, 
$\parallel$ and $\perp$ subscripts are used to
distinguish directions parallel and perpendicular to the drive. 
In equation (\ref{parallel}),
  $\partial_\parallel$ with a prime  acts on the $x'$ coordinate. 
${\cal H}$
is the usual Ginzburg-Landau Hamiltonian for the lattice-gas
\begin{eqnarray}
{\cal H}=\int d^dx\{\frac{1}{2}(\nabla S)^2+
\frac{1}{2}\tau S^2+\frac{u}{4!}S^4\}, \label{ginzburg}
\end{eqnarray} 
where $\tau \propto (T-T_c)$ measures the deviation from the critical
temperature.  Although the  hamiltonian is isotropic,
the drive induces  anisotropic $\tau$'s. This is usually taken care 
of by introducing two $\tau$'s, $\tau_\parallel$ and $\tau_\perp$, 
which behave differently as $T$ approaches 
$T_c$ \cite{leung}.  We, therefore, need to consider two
such $\tau$'s   in our equation of motion.                 
  Note that we have incorporated the non-local
conservation in the parallel component of the current in equation
(\ref{parallel}).  The local conservation corresponds to $\chi({\bf
  x}-{\bf x}')=\delta({\bf x}-{\bf x}')$.  Since the effect of the
drive is expected to be predominant along its direction, we introduce
the non-locality only in the parallel direction.  Generalization of this,
however, is possible along the same line.  
When the drive is switched on, there is an additional particle current
along the drive. This leads to an additive term proportional to $g
\partial_\parallel S^2$ in equation (\ref{cons}) with $g$ being related to
the strength of the drive.  Equations
(\ref{cons})-(\ref{ginzburg}), along with this nonlinear 
term due to the drive lead to the final Langevin equation which 
we have written explicitly in the appendix. 
 
Different forms of
$\chi({\bf x}-{\bf x'})$ can be considered in 
general.  Since any short-range
$\chi$ under renormalization would look like a delta-function, in the
large scale limit, $\chi$ is going to matter only if it is
long-ranged.  We, therefore, consider a special case where
\begin{eqnarray}
\chi({\bf x}-{\bf x}')\sim \delta({\bf x}_\perp-{\bf x}'_\perp)
(x-x')_\parallel^{\sigma-1}.
\end{eqnarray}
In particular, in the Fourier space, 
$\chi(k)=\rho_{nl} k_\parallel^{-\sigma}$
with $\rho_{nl}$ denoting 
 the strength of the non-local conservation.
The local conservation corresponds to $\sigma=0$, and $\sigma=2$ would
imply a global conservation in the parallel direction. 
Since the inter-particle interaction is expected to remain unaffected
by the drive in the transverse direction, we expect $\tau_\perp$ to 
vanish as the critical temperature is approached. 
Considering a positive finite value for $\tau_\parallel$, 
 it can be seen that the leading momentum
dependent terms in the equation of motion are $k_\perp^4$ and
$k_\parallel^{2-\sigma}$.  This leads to an anisotropic scaling
\begin{eqnarray}
k_\parallel\sim k_\perp^{1+\Delta}
\end{eqnarray}
with $\Delta=(2+\sigma)/(2-\sigma)$. Therefore, in the presence of 
a non-local conservation, the  anisotropy 
 is more pronounced 
than the corresponding local case.  
%The ansiotropy is even stronger than the usual 
%conservation case as long as $0<\sigma<2$. 
In the large length scale limit, all the irrelevant terms drop 
out \cite{ziarev} and
in the Fourier space, we have the following equation of motion 
\begin{eqnarray}
[ i\omega+\lambda k_\perp^4+\lambda\chi(k) k_\parallel^2] 
S({\bf k},\omega)
&&-\frac{1}{2}\lambda g ik_\parallel \int d{\bf k}_1 
d\omega_1 S({\bf k}_1,\omega_1) 
S({\bf k}-{\bf k}_1,\omega-\omega_1)\nonumber\\
&&=-i {\bf k}_\perp\cdot {\bf \zeta}_\perp({\bf k},\omega). \label{langvin2} 
\end{eqnarray}
In the Fourier space,  the 
spatially and temporally uncorrelated noise 
 is described by the second moment
\begin{eqnarray}
\langle i {\bf k}_\perp\cdot {\bf \zeta}_\perp({\bf k},\omega)
i{\bf k}'_\perp\cdot {\bf \zeta}_\perp({\bf k'},\omega') \rangle=
2 D k_\perp^2 \delta({\bf k}+{\bf k'})\delta(\omega+\omega'),
\end{eqnarray}    
with $D$ characterizing the noise amplitude.
In equation (\ref{langvin2}), we have  dropped a naively irrelevant term 
$\frac{u}{3!} \nabla_\perp^2 S^3$ which assures 
stability below the criticality. This term, identified as 
a dangerously irrelevant term, plays a crucial role in determining the 
equation of state and the critical exponent $\beta$ for the KLS model
\cite{janssen}.

It can be  very easily seen from the real-space version of equation 
(\ref{langvin2}), that it possesses Galilean invariance \cite{janssen}
 under the transformation
\begin{eqnarray}
S({\bf x},t) \rightarrow S({\bf x}+\lambda g a {\bf e}t,t)+a,\label{gal1}
\end{eqnarray}
where $a$ is a small continuous parameter and ${\bf e}$ is the 
unit vector in the direction of the drive. 
Under this transformation, the stochastic equation ({\ref{langvin2})
is subjected to a noise $\nabla_\perp \cdot {\bf \zeta}'_\perp({\bf x}',t)=
\nabla_\perp\cdot 
{\bf \zeta}_\perp({\bf x}+\lambda g a {\bf e} t, t)$. 
The noise correlation is modified as 
\begin{eqnarray}
\langle {\bf \nabla}_\perp\cdot {\bf \zeta}'_\perp({\bf x}'_1,t_1) 
{\bf \nabla}_\perp\cdot {\bf \zeta}'_\perp({\bf x}'_2,t_2)\rangle &&=
\langle {\bf \nabla}_\perp\cdot {\bf \zeta}_\perp({\bf x}_{1\perp},
{ x}_{1\parallel}+\lambda a g t_1,t_1)
{\bf \nabla}_\perp\cdot {\bf \zeta}_\perp({\bf x}_{2\perp},x_{2\parallel}+
\lambda a g t_2,t_2)\rangle
\nonumber\\
&&={\bf \nabla}_\perp^2  F({\bf x}_{1\perp}-
{\bf x}_{2\perp},x_{1\parallel}-x_{2\parallel}+
\lambda a g (t_1-t_2), t_1-t_2),
\end{eqnarray}
where, for generality, we have retained an arbitrary function 
$F$ that represents the noise correlation in the original equation.
In the case of a $\delta$-correlation in time, 
$F(x_\perp, x_\parallel,t)=\delta(t) f(x_\perp, x_\parallel)$, 
the correlation of the new 
noise is the same as the original one. This, however, is not true 
when there is a long-range temporal correlation in the noise.
Therefore the stochastic equation (\ref{langvin2}) is invariant 
under Galilean  transformation                               
only when the noise has a short-range correlation in time.

For local conservation, $\sigma=0$, a dimensional analysis  
leads to
$\Delta=1$.  Because of the anisotropy, we need to define two $\nu$
like exponents associated with the correlation lengths in the
transverse and longitudinal directions. In the same way, one can
define two $z$ like dynamic critical exponents and different
$\eta$-like anomalous exponents associated with the structure factor.
The detailed relationship of these exponents with the original Ising
exponents $z$ or $\eta$ can be worked out \cite{ziarev}. A  one-loop RG
analysis further reveals that there exists a nontrivial infra-red
stable fixed point \cite{janssen} 
for the drive that determines the critical behavior
of the system below the upper-critical dimension $d_c=5$.  Universal
exponents acquire corrections in the $\epsilon\ (=5-d)$-expansion
scheme and the independent exponents are
\begin{eqnarray}
\Delta=1+\frac{\epsilon}{3}, \  z=4, \ \eta=0,\ \nu=1/2,\ \beta=1/2. 
\end{eqnarray}
For $d>5$, the drive is irrelevant and the  critical exponents 
are  mean-field like.

%\section{Scaling}
\subsection{Renormalization Group analysis}
Since the RG analysis in the following does not lead to any
renormalization of $\lambda$, we set $\lambda=1$.
However, for book-keeping purposes,
it is convenient to maintain explicitly a coefficient
 of $k_\perp^4$ term in the square
bracket of equation (\ref{langvin2}).
This coefficient, henceforth, is
denoted by $\overline\nu$.
In the  RG analysis, we look for the scale invariance of
all the  coefficients.   
                              
A naive dimensional analysis   
can be done for the present system.  We
observe how the equation of motion changes under a change 
of scale $x_\perp\rightarrow b x_\perp, x_\parallel \rightarrow b^{1+\Delta}
x_\parallel$ and $t\rightarrow b^z t$, where $z$ is the dynamic
exponent.  Assuming that the field scales as $S\rightarrow b^\chi S$
under this transformation, we find that various parameters scale as
\begin{eqnarray}
\overline\nu\rightarrow b^{z-4} \overline\nu,\\
\rho_{nl} \rightarrow b^{z+(1+\Delta)(\sigma-2)} \rho_{nl},\\
g\rightarrow b^{\chi-1-\Delta+z} g,\\
D\rightarrow b^{-2-(d+\Delta+z)-2\chi+2 z} D.
\end{eqnarray}
For local conservation, $\sigma=0$, and for $g=0$, the invariance of
$\overline\nu, \rho_{nl}$ and $D$ leads to
 $z=4,\ \Delta=1$ and $\chi=(1-d)/2$.
These exponents are associated with the trivial fixed point $g=g^*=0$.
Around this fixed point, the drive strength $g$ scales as $b^{(5-d)/2}
g$ and this implies that the upper-critical dimension is $5$.  For
$\sigma>0$, $\rho_{nl}$ is relevant around the trivial fixed point and
we perform an RG analysis in the following to study its impact on the
known one-loop results.

Many technicalities related to the one-loop dynamic RG analysis done
in the following are similar to \cite{medina} even though there is no
immediate connection between the two problems.  We, therefore, skip
the details and mention the steps for sake of completeness. 
A convenient 
starting point is to rewrite equation (\ref{langvin2}) as 
\begin{eqnarray}
S({\bf k},\omega)=- i {\bf k}_\perp\cdot \zeta_\perp({\bf k},\omega) 
G_0({\bf k},\omega)+&&\frac{1}{2} g G_0({\bf k},\omega)
(i k_\parallel)\times\nonumber\\&& \int d{\bf k}_1 d\omega_1
S({\bf k}_1,\omega_1) S({\bf k}-{\bf k}_1,\omega-\omega_1),\label{perturb0}
\end{eqnarray}
with a bare propagator 
\begin{eqnarray}
G_0({\bf k},\omega)=(i\omega+\overline\nu k_\perp^4+
\rho_{nl} k_\parallel^{2-\sigma})^{-1}. \label{propagatorb}
\end{eqnarray}
The bare vertex is given by $\frac{1}{2} g (i k_\parallel)$. 
Diagrammatic representations of this equation 
and the vertex are shown in Fig 1. 
\begin{figure}[htbp]
  \begin{center}
%    \narrowtext
%   \psfig{file=nonloc_fig0.eps,width=3in,angle=0}
   \includegraphics[width=5in,clip]{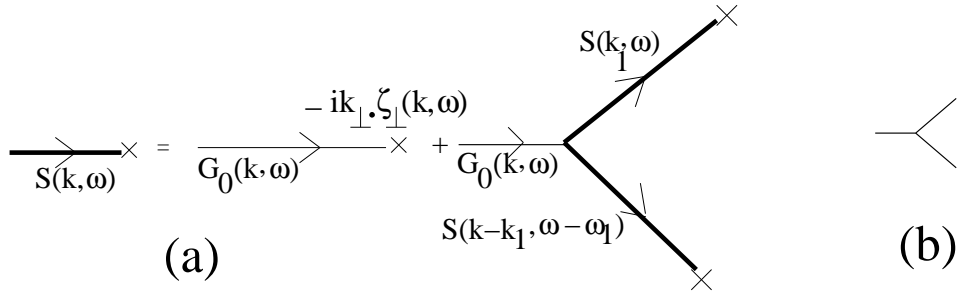}
    \caption{(a) Diagrammatic representation of equation 
(\ref{perturb0}). (b) Three point vertex. $\times$ represents the noise 
and a thin line with an arrow represents a bare propagator. }
\label{fig:0}
  \end{center}
\end{figure}

The normal procedure is to perform an iterative expansion where $S$ in
equation (\ref{perturb0}) is replaced by itself.  This is continued up
to the desired order and different 
terms in the series are, then, averaged over the noise. 
 Various terms in the perturbative series, thus obtained,
give rise to effective average quantities.  We find that the 
 effective propagator 
$G({\bf k},\omega)$ defined by $S({\bf k},\omega)= 
-i{\bf k}_\perp\cdot {\bf \zeta}({\bf k},\omega) G({\bf k},\omega)$ is
\begin{eqnarray}
&&G({\bf k},\omega)=G_0({\bf k},\omega)-2(\frac{1}{2} g)^2
2D k_z G_0^2({\bf k},\omega) \{\int \frac{d{\bf k}_1
d\omega_1}{(2\pi)^{d+1}} (k_{1z}+
\frac{1}{2} k_z) ({\bf k}/2-{\bf k}_1)_\perp^2\times\nonumber\\
&& G_0({\bf k}_1+{\bf k}/2,\omega_1+\omega/2)
\mid G_0({\bf k}_1-{\bf k}/2,\omega_1-\omega/2)\mid^2+
\int \frac{d{\bf k}_1 d\omega_1}{(2\pi)^{d+1}}
(k_z/2-k_{1z})\times\nonumber\\&&
({\bf k}_1+{\bf k}/2)_\perp^2 G_0({\bf k}/2-{\bf k}_1,\omega/2-\omega_1)
\mid G_0({\bf k}_1+{\bf k}/2,\omega_1+\omega/2)\mid^2\}.\label{propagator}
\end{eqnarray}                                                                  
 A diagrammatic representation of the terms
appearing in this equation  is given in Fig. 2(a).
\begin{figure}[htbp]
  \begin{center}
%    \narrowtext
%   \psfig{file=nonloc_fig1.eps,width=3in,angle=0}
   \includegraphics[width=5in,clip]{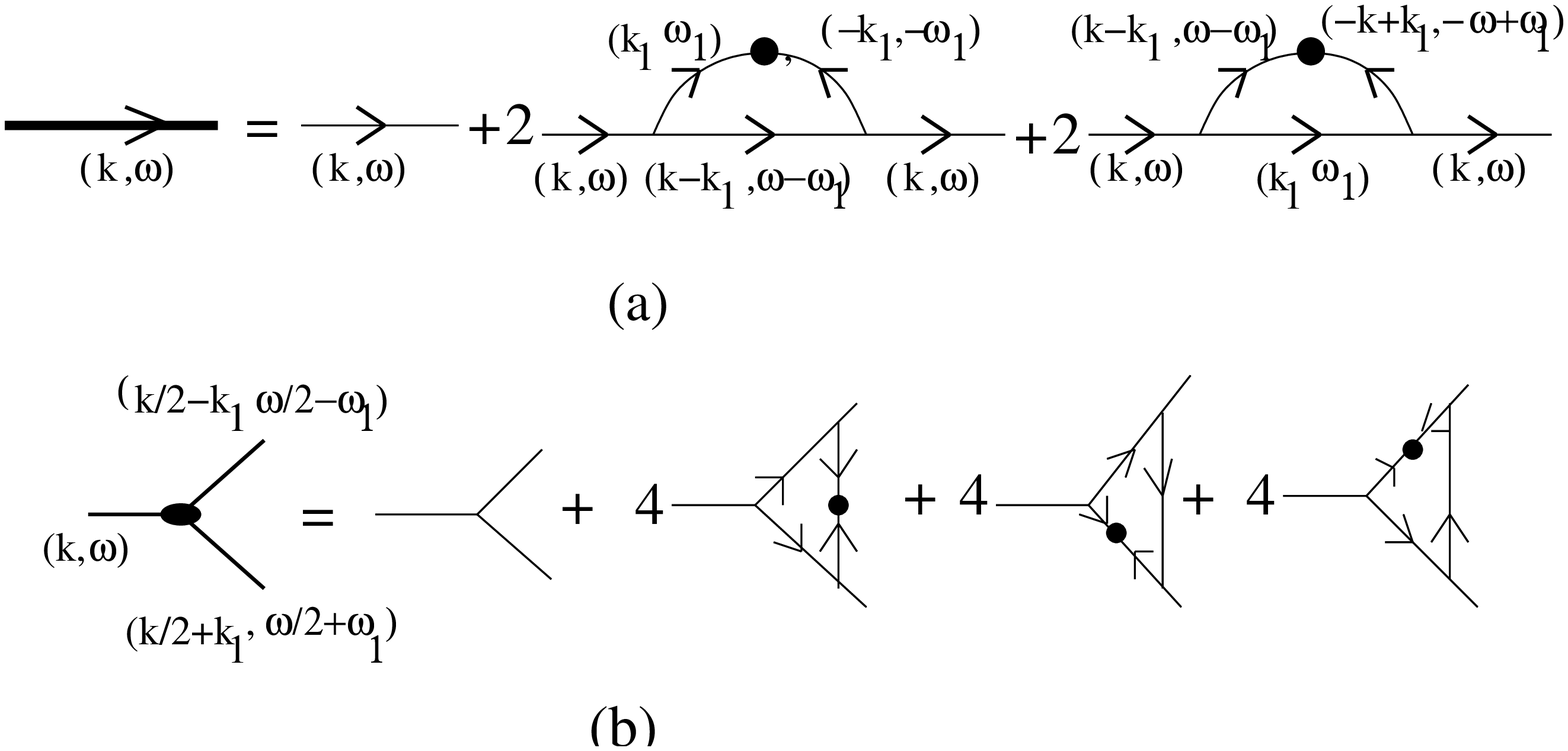}
    \caption{ (a) Diagrams contributing to the renormalization of the
propagator. (b) Diagrams contributing to the renormalization of the
vertex. On the left hand sides, we have the effective propagator (in (a))
and the effective vertex (in (b)). Filled dots represent noise contraction.
Numbers in front of the diagrams are the  combinatorial factors that
originate from possible noise contractions.}
    \label{fig:1}
  \end{center}
\end{figure}                                                                   
The one-loop corrections obtained from the perturbative series 
have divergences in
the $k_1 \rightarrow 0$ limit.  The renormalization group amounts to
avoiding these singularities by re-summation. There is also 
an ultra-violet cutoff $\Lambda$ in the momentum and this is due to the
existence of a short distance cutoff in the system. 
We use the momentum shell
renormalization scheme, where fluctuations over the momentum shell
$\Lambda e^{-l}<k_1<\Lambda$ are integrated out and subsequently a
rescaling in the momentum as $k_1\rightarrow k_1 e^{-l}$ is done to 
retrieve the original cutoff $\Lambda$.  
The rescaling of momentum is
the same as that carried out earlier with $b=e^{l}$.

 An effective noise amplitude $\tilde D$
can be defined in a similar way as
\begin{eqnarray}                                                    
\langle S^*({\bf k},\omega) S({\bf k},\omega)\rangle= 
2k_\perp^2 \tilde D G({\bf k},\omega) G({\bf -k},\omega).
\end{eqnarray}                                                    
To see the renormalization of the drive, one has to calculate the 
effective    three-point vertex function $\Gamma$, where the bare vertex 
function is given by  $\Gamma_0=i k_\parallel (g/2)$.
However, it can be checked that no one-loop correction to the effective 
noise amplitude is present here. This is essentially due to the 
fact that the nonlinear term due to the drive 
 is accompanied by an overall factor $k_\parallel$.  
The drive term also
does not require any 
 renormalization and this is due to the Galilean invariance
which is valid in this case as well.  RG transformation, being analytic in
nature, cannot renormalize $\rho_{nl}$.

It is easy to check that starting with the  propagator in equation
(\ref{propagatorb}), a term 
proportional to $k_\parallel^2$ is always generated. We, therefore, 
incorporate this term, with a coefficient $\rho_l$, 
in the bare propagator $G_0({\bf k},\omega)$ which   now appears as 
\begin{eqnarray}
G_0({\bf k},\omega)=(i\omega+\overline \nu k_\perp^4+
\rho_{nl} k_\parallel^{2-\sigma}+\rho_l k_\parallel^2)^{-1}.
\end{eqnarray}
In the hydrodynamic limit, $({\bf k},\omega)\rightarrow 0$, 
the  effective $\rho_l$ is given by 
\begin{eqnarray}
\rho^{\rm eff}_l=\rho_l^0+(g/2)^2[A+B],
\end{eqnarray}
where 
\begin{eqnarray}
A=\int \frac{d{\bf k}_1 d\omega_1}{(2\pi)^{d+1}}
 k_{1\perp}^2 G_0(-k_1,-\omega_1)^2 G_0(k_1,\omega_1)
\end{eqnarray}
and 
\begin{eqnarray} 
B=\int \frac{d{\bf k}_1 d\omega_1}{(2\pi)^{d+1}}
 k_{1\perp}^2 G_0(-k_1,-\omega_1) 
G_0(k_1,\omega_1)^2.
\end{eqnarray}
To do the momentum shell integration, we
take care of the anisotropy in the transverse and longitudinal directions 
by  choosing  polar coordinates \cite{aharony}
\begin{eqnarray}
\overline\nu^{1/2}k_{1\perp}^2
=r \sin\theta,\ \  \rho_l k_{1\parallel}^2=r^2 \cos^2\theta.
\end{eqnarray} 
The momentum integration over the shell with 
inner and outer radii  
$\Lambda e^{-l}$ and $\Lambda$ with infinitesimal $l$ leads to 
\begin{eqnarray}
\rho^{\rm eff}_l=\rho^o_l+\frac{g^2 D}{8 {\rho^o}_l^{1/2}
\overline\nu^{(d+1)/4}}
l \frac{S_{d-1}}{(2\pi)^d} I(d,\sigma,\rho_r), \label{rholoneloop}
\end{eqnarray}
where 
\begin{eqnarray}
I(d,\sigma,\rho_r)=\int d\theta \frac{\sin^{(d-1)/2}\theta}
{(1+\rho_r \cos^{2-\sigma}\theta)^2},\\
{\rm and}\ \ \ \ \ \ \ \  \rho_r=\rho_{nl}/\rho_l^{(2-\sigma)/2}.
\end{eqnarray}
$S_{d-1}$ represents the surface area of a unit $d-1$ dimensional 
sphere. In equation (\ref{rholoneloop}), we have used $\Lambda=1$. 

After rescaling to retrieve
the original momentum cutoff, we have the following RG equations,
describing the flow of the parameters under the change of the length
scale
\begin{eqnarray}
&&\frac{d\overline\nu }{dl}=(z-4)\overline\nu, \label{nurg}\\
&&\frac{d\rho_{nl}}{dl}=[z-(2-\sigma) (1+\Delta)]
\rho_{nl},\label{rhonlrg}\\
&&\frac{d\rho_l}{dl}=\rho_l[(z-2(1+\Delta))+\frac{u D}{8} 
\frac{K_{d-1}}{2\pi}
I(d,\sigma,\rho_r)], \label{rholrg}\\
&&\frac{dD}{dl}=(z-2\chi-d-2-\Delta)D,\label{noiserg}\\
&&\frac{du}{dl}=[2\chi+(1+\Delta)+z/2-\frac{1}{4}(d+1)(z-4)] 
u-\nonumber\\ && \frac{3}{16} u^2 D\frac{K_{d-1}}{2\pi}
 I(d,\sigma,\rho_r),\label{driverg}
\end{eqnarray}
where $u=g^2\rho_l^{-3/2}\overline\nu^{-\frac{1}{4}(d+1)}$ and  
$K_{d-1}=S_{d-1}/(2\pi)^{d-1}$.

An RG  equation for  $\rho_r$ can be obtained from equations 
(\ref{rhonlrg}) and 
(\ref{rholrg}).  
\begin{eqnarray}
\frac{d\rho_r}{dl}=\rho_r[\frac{\sigma z}{2}-
(2-\sigma)\frac{u D}{16} \frac{S_{d-1}}{(2\pi)^d}
 I(d,\sigma,\rho_r)].\label{rhor}
\end{eqnarray}
The trivial fixed point $\rho_r^*=0$ corresponds to the local
conservation.  In that case, there exists a non-Gaussian infra-red
stable fixed point $ u^* =32 \epsilon/3 C_1$ where
$C_1=\frac{S_{d-1}}{(2\pi)^d}\sqrt{\pi}
\Gamma[(1+d)/4]/\Gamma[(3+d)/4$.  Further, 
the scale invariance of $\overline\nu$ and
$\rho_l$  implies $z=4$ and $\Delta=1+\epsilon/3$ respectively.
The invariance of $D$ under RG transformation leads to
$\chi=(2-d-\Delta)/2$. Substituting the expression of $\Delta$, we
have $\chi=\frac{1}{2}(1-d-\epsilon/3)$.  The exponent $\chi$
describes the scaling of the correlation function $\langle
\phi(x,t) \phi(0,0)\rangle$.  In the presence of the anisotropic
scaling, this correlation function scales as
\begin{eqnarray}
\langle\phi(x,t) \phi(0,0)\rangle \sim x_\perp^{2\chi} 
f_1(x_\parallel/x_\perp^{1+\Delta},t/x_\perp^z),\label{correlation}
\end{eqnarray}                    
where $f_1$ is an appropriate scaling function.                  
A nontrivial fixed point for $\rho_r$ exists if  
\begin{eqnarray}
\frac{\sigma z}{2}=(2-\sigma) \frac{u D}{16} \frac{K_{d-1}}{2\pi}
 I(d,\sigma,\rho_r).
\end{eqnarray}
Since the nontrivial fixed point for $u$ is  
\begin{eqnarray}
u^*=(3 z/2-1-d) \frac{16}{3 D I(d,\sigma,\rho_r) K_{d-1}/(2\pi)},
\label{fixedpoint}                                     
\end{eqnarray}  the exponent relation
\begin{eqnarray}
\frac{\sigma z}{2}=\frac{(2-\sigma)}{3} (3 z/2-1-d)\label{exprelation}
\end{eqnarray}
is satisfied whenever $u=u^*$ and $d\rho_r/dl=0$.  With $z=4$, one
finds a critical value of $\sigma$, $\sigma_c=
\frac{2\epsilon}{\epsilon+6}$ from equation (\ref{exprelation}).  At
this critical value, $\rho_r$ is marginal and the scale invariance of
$\rho_{nl}$ in equation (\ref{rhonlrg}), yields $\Delta=1+\epsilon/3$.
Obviously, this value of $\Delta$ also leads to the invariance of
$\rho_l$. For $\sigma<\sigma_c$, $\rho_r$ is irrelevant and flows to
zero in the asymptotic limit. The fixed point in equation
(\ref{fixedpoint}) and the anisotropy exponent are the same as those
in the local conservation case.  For $\sigma>\sigma_c$, $\rho_r$ is
relevant and one needs to extend equation (\ref{rhor}) beyond one-loop to
get a fixed point.  If there exists a fixed point, invariance of
$\rho_{nl}$ would lead to $\Delta=(2+\sigma)/(2-\sigma)$.  A numerical
simulation is likely to reveal the universal properties for large
$\sigma$.

 It may be noted that in many cases the approach to the short-range 
limit is somewhat subtle.  
For ferromagnets with long-range exchange 
of the  form $r^{-(d+\sigma_e)}$, with 
 $r$ as the distance between two spins,        
critical exponents  show  apparent discontinuity as 
the short-range limit $\sigma_e\rightarrow 2$ \cite{fisher} is approached.
Later, this problem has been 
 sorted out  by Sak \cite{sak}, in his momentum-shell RG analysis, 
by taking the corresponding local
  term that is generated under renormalization. Although irrelevant, 
this local term  becomes at least 
marginal as $\sigma_e\rightarrow 2$ and competes 
with the non-local term. This interplay leads to the fact that 
as $\sigma_e$ approaches  $2$ from below,  short-range critical 
exponents are found if $\sigma_e> \sigma_{ec}$ where $\sigma_{ec}$ is
determined by  short-range exponents.          
In the field theoretic RG formulation, the apparent discontinuity 
is  removed  by 
performing a double expansion in the  corresponding 
$\epsilon$$(=d_c-d)$ and the 
 deviation of $\sigma_e$ from its short-range value \cite{honkonen}. 
As the short-range limit is approached, this deviation is small 
and is treated on equal footing with $\epsilon$. The problem of 
driven diffusive system with a long-range spatially correlated noise 
\cite{js}   is 
 more intricate due to the existence of two fixed points 
in the short-range case and only one fixed point in the corresponding 
long-range version. In our case, the  
discontinuity related problem 
is naturally taken care of by the local term that is 
generated.  In  a similar way as \cite{sak}, the short-range behavior 
is retrieved for $\sigma<\sigma_c$, where
$\sigma_c$ is obtained from  the short-range exponents.
                         
A more general form of the non-local conservation can exhibit what
happens in the case of completely global conservation. This
generalization requires a reconsideration of the scaling and relevance
or irrelevance of all the terms  originally present in the model.
In our case, for the global conservation limit, $\sigma
\rightarrow 2$, one would require such consideration since there is no
$k_\parallel$ dependent term in the free part of equation
(\ref{langvin2}).

% one  has a critical 
%value for $\sigma$  at which corresponds to fixed points for 
%$u$ and $\rho_r$.
%\begin{figure}[htbp]
%\begin{center}
%\narrowtext
%\psfig{file=nonloc_fig1.eps,width=4in,angle=0}
%\label{}
%\end{center}
%\end{figure}

%1.     Around the trivial(Gaussian) fixed point where 
%$z=4$ and $a=2$, one has 
%$\rho_{nl}$ relevant 
%for positive $\sigma$. That is likely to make $I(d,\sigma,\rho_r)->0$. 
%
%Does it 
%trivially 
%mean that effect of drive vanishes asymptotically?
%2.     The other possibility is that the anisotropy 
%exponent $a$ changes from 
%$2$ to 
%$a=z/(2-\sigma$. This is quite natural from the naïve 
%scaling of the differential 
%eqn. This makes $(z-2 a)=-\sigma z/(2-\sigma)$ 
%negative for $+$-ve $\sigma$ and 
%$(z-2 a$ +ve for $-$ ve $\sigma$.  In the first case $I(d,\sigma,\rho_r)$ 
%should vanish, 
%where as in the second case $I(d,\sigma,\rho_r)$ 
%should lead to finite value.

\section{Temporally correlated noise}
In this section, we consider a noise with  a long-range temporal
correlation.   
 A large simplification in the RG analysis follows from the
invariance of the system under Galilean transformation described in 
equation (\ref{gal1}).                                   
The RG transformation, that preserves this symmetry,
assures that the drive is not renormalized in the process.  With a
long-range temporal correlation in the noise, the Galilean invariance
is lost  and the flow equation associated with the drive
acquires  contributions from  one-loop terms in the perturbation
series. 
%\footnote{Instead of temporal correlation, spatially
%  correlated noise can also be studied using the same procedure.
%  However, we do not expect a renormalization of the drive since
%  Galilean invariance is not disturbed in this case. Such long-range
%  spatial correlation is expected to change only the scaling
%  dimensions of the parameters.}  
To obtain this, we, here, proceed with the noise correlation
\begin{eqnarray}
\langle i {\bf k}_\perp\cdot {\bf \zeta}_\perp(k,\omega)
i{\bf k}'_\perp\cdot {\bf \zeta}_\perp(k',\omega') \rangle=
2 D k_\perp^2 \omega^{-2\theta_d}\delta({\bf k}+{\bf k'})
\delta(\omega+\omega').
\end{eqnarray}
For   $\theta_d=0$, all the known results are expected to follow.

The renormalization of $\rho_l$ follows from the same expansion done
for equation (\ref{perturb0}), except for a new factor
$\mid\omega\mid^{-2\theta_d}$ associated with the averaging over the 
noise in the present case.  
The three point vertex function now
 acquires contributions from one-loop terms. 
Three diagrams that contribute to the renormalized vertex at the
one-loop level are shown in Fig.2(b).  Evaluating these diagrams, one
finds the effective drive 
\begin{eqnarray}
&&g_{\rm eff}= g[1+4(\frac{1}{2} g)^2 2D 
\int \frac{d{\bf k}_1 d\omega_1}{(2\pi)^{d+1}} k_{1\parallel}^2 k_{1\perp}^2 
G_0({\bf k}_1,\omega_1)^2 \times\nonumber\\&& 
G(-{\bf k}_1,-\omega_1)^2\mid \omega_1\mid^{-2\theta_d}-
  8(\frac{1}{2} g)^2 2D\times\nonumber\\ &&
\int \frac{d{\bf k}_1 d\omega_1}{(2\pi)^{d+1}} 
k_{1\parallel}^2 k_{1\perp}^2 \mid \omega_1\mid^{-2\theta_d}
G_0({\bf k}_1,\omega_1)^2 \mid G({\bf k}_1,\omega_1)\mid^2].
\end{eqnarray}
After a straight forward evaluation of the one-loop contribution over
the momentum shell and subsequent rescaling of the momentum, we obtain
the flow equation for the drive
\begin{eqnarray}
&&\frac{dg}{dl}=g\{(z+\chi-1-\Delta)-\nonumber\\ &&\frac{K_{d-1}}{2\pi} 
g^2 D \rho_l^{-3/2} {\overline{\nu}}^{-1/4(d+1)} B\int_{-\infty}^{\infty}
\frac{d\omega}{(2\pi)} 
\frac{(1-3 \omega^2)}{(1+\omega^2)^3}\mid\omega\mid^{-2\theta_d}\},
\end{eqnarray}
where $B=\int_0^{\pi/2}d\theta \sin^{(d-1)/2}\theta \cos^2\theta
=\frac{\sqrt \pi}{4} \Gamma[(1+d)/4]/\Gamma[(7+d)/4]$.  As expected,
for $\theta_d=0$, the above integral over $\omega$ vanishes.  In terms
of $u$, we have the RG flow equations
\begin{eqnarray}
&&\frac{d{\overline{\nu}}}{dl}={\overline\nu}(z-4),\\
&&\frac{dD}{dl}=[z-2\chi-2-d-\Delta+2 z\theta_d]D,\\
&&\frac{d\rho_l}{dl}=\rho_l[(z-2 -2\Delta)+
u \frac{D}{8} \frac{K_{d-1}}{2\pi}A(1+2\theta_d) \sec(\pi \theta_d)],\\
&& \frac{d u}{dl}=u[\frac{3 z}{2}+2 z \theta_d-(d+1)-
\frac{1}{4}(d+1)(z-4)]\nonumber\\
&&-u^2 A (\theta_d B/A-\frac{3}{16}) 
D\frac{K_{d-1}}{2\pi}(1+2\theta_d) \sec\pi\theta_d, \label{urg}
\end{eqnarray}
where 
$A=\int_0^{\pi/2} d\theta \sin^{(d-1)/2}\theta=\frac{\sqrt{\pi}}{2}
\Gamma[(1+d)/4]/\Gamma[(3+d)/4]$.
  As before, the noise amplitude 
$D$ remains un-renormalized here.  
 In the case of  short-range temporal correlation, the only quantity 
that requires renormalization is $\rho_l$. Although the drive is not 
renormalized there  due to the Galilean invariance, 
the effective drive dependent parameter $u$ acquires a one-loop like term 
as a consequence of its definition. 
As a result  of this interplay between 
$\rho_l$ and $u$ and 
the Galilean invariance, 
the anisotropy exponent becomes exact. In our case, the drive is 
renormalized independently and, 
therefore, one-loop results are not  exact anymore.

Scale invariance of ${\overline\nu}$ 
and $D$ leads to
 \begin{eqnarray}
z=4, \ {\rm and}\ 
\chi=\frac{1}{2}(2-d-\Delta+8\theta_d).
\end{eqnarray}                                         
A $\theta_d$-dependent fixed point for $u$ follows from equation
(\ref{urg}). At this fixed point
\begin{eqnarray}
\frac{d\rho_l}{dl}=\rho_l[z-2-2\Delta-
\frac{\epsilon_\theta}{8} \frac{1}{(\theta_d B/A-3/16)}], 
\end{eqnarray} 
where $\epsilon_\theta=5-d+8\theta_d$.  The scale invariance of $\rho_l$
now leads to a $\theta_d$-dependent anisotropy exponent
\begin{eqnarray}
\Delta=1-\frac{\epsilon_\theta}{16(2\theta_d/(3+d)-3/16)}.
\end{eqnarray}
Using this value of the anisotropy exponent at the nontrivial fixed
point, we have a $\theta_d$-dependent $\chi$ that describes the
scaling of the correlation function in equation (\ref{correlation}) in
the presence of a temporally correlated noise. All the universal 
exponents go over to the corresponding short-range
values continuously as $\theta_d\rightarrow 0$.                                            

\section{Conclusion}

To summarize our results, we have studied the effect of weakening of two
basic ingredients, namely the particle conservation and the Galilean
invariance,  responsible for  nontrivial scaling in driven
diffusive systems.
We have done  this by introducing a non-locality in the
conservation of particles and a temporally correlated noise. 
 With a long-range temporal
correlation of the noise, the exactness of the previously known
one-loop results is destroyed due to the breakdown of the
Galilean invariance. We find a new anisotropic exponent that
depends on the power of the long-range correlation. The scaling of the
two-point correlation with the change of the length scale is also
found out.
For a non-local particle  conservation,  we show how the
one-loop RG calculation is modified 
by considering a simple modification of the KLS model.
The non-locality is introduced through  a non-local kernel 
in the parallel component of the Langevin current. 
This procedure can be extended to study the effect of more 
general forms of non-local conservation.
 Our dynamic RG analysis shows that the
non-locality is irrelevant below a critical power $\sigma_c$ and is
marginal for $\sigma=\sigma_c$.  Above $\sigma_c$, the non-locality is
relevant.  The relevance of the non-locality deserves a more detailed
investigation since there exists a possibility of a crossover to a
regime with different universal exponents.  Besides 
the theoretical interest, non-local dynamics has its own relevance 
due to speeding up of  numerical simulations implemented by different 
kinds of non-local moves \cite{klein}. We expect that incorporating 
various non-local features in the  simulations of driven systems
would be interesting and also would be a crucial check of the theory
employed here.

\appendix
\section{The full Langevin equation}
The full Langevin equation that follows from equations 
(\ref{cons})-(\ref{ginzburg}) in a straight forward way is   
\begin{eqnarray}
&&\partial_t S({\bf x},t)=
\lambda \nabla_\perp^2(\tau_\perp S-\nabla_\perp^2 S)
+\lambda \frac{u}{3!} \nabla_\perp^2 S^3+\nonumber+
\lambda \partial_\parallel\int d^dx'\chi({\bf x}-
{\bf x'})\partial_\parallel'
(\tau_\parallel S'-\partial_\parallel'^2 S')+\nonumber\\&& 
+\lambda \kappa\frac{u}{3!}
\partial_\parallel \int d^dx' \chi({\bf x}-{\bf x'})\partial_\parallel' S'^3-
\lambda \nabla_\perp^2\partial_\parallel^2 S-\lambda \partial_\parallel 
\int d^dx' \chi({\bf x}-{\bf x'}) 
\partial_\parallel'(\nabla'^2 S')+g \partial_\parallel S^2
-\nonumber\\&&
\nabla_\perp\cdot \zeta_\perp({\bf x}_\perp,t)-
\partial_\parallel \int d^dx' 
\chi({\bf x}-{\bf x'}) \zeta_\parallel ({\bf x'},t).
\end{eqnarray}}
Unlike $S$, $S'$ is a function of ${\bf x}'$ coordinate. 
A scaling analysis similar to that done for the usual KLS model 
($\sigma=0$)  near criticality \cite{ziarev}
 implies that terms that are irrelevant
in the local case remain 
irrelevant in the presence of a more  pronounced anistropy for 
the non-local case ($\sigma\neq 0$). 

{\bf Acknowledgment}
Financial support from Indian Institute of Technology, Kanpur 
is acknowledged.

%\end{multicols}
\end{document}